\documentclass[11pt,oneside,a4paper]{article}

\usepackage{latexsym}
\usepackage{amssymb, amsmath}
\usepackage{exscale}
\usepackage{bm, graphics}
\usepackage{graphicx, subfigure, pstricks}

\newcommand{\D}{\mbox{\rm d}}

\newcommand{\uh}[1]{\underline{\hat{#1}}}
\newcommand{\ket}[1]{$\left|#1\right\rangle$}

\title{Generation of time-bin entangled photon pairs using 
a single three-level emitter
}

\author{M. Khanbekyan \\ 
	Institut f{\"u}r Physik,\\
      Otto-von-Guericke-Universit{\"a}t Magdeburg, Magdeburg, Germany\\
        khanbekyan@gmail.com
}

\begin{document}
\maketitle
\begin{abstract}
	% enabling complete control of a time-bin qubit without the use of an 
	%interferometer.
	We study single-photon emission of a classically pumped three-level 
	$\Lambda$-type emitter in a high-$Q$ cavity. In particular, generation of 
	single-photon time-bin double-peak wave packets is shown.
\end{abstract}

%%%%%%%%%%%%%%%%%%%%%%%%%%%%%%%%%%%%%%%%%%%%%%%%%%%%%%%%%%%%%%%%%%%%%%%

\section{Introduction}
\label{introduction}

A source of entangled photon pairs is an essential element for quantum 
communication~\cite{gisin:165}, linear optics quantum computing~\cite{knill:46} 
and processing of quantum information protocols such as quantum 
teleportation~\cite{bouwmeester:575, riedmatten:050302} and quantum 
cryptography~\cite{jennewein:4729,deng:042317}. Commonly, the employed entangled
states in experiments are polarization-entangled photons 
generated by e.g. spontaneous parametric down-conversion~\cite{kwiat:4337} or 
biexciton--exciton cascade of single semiconductor quantum 
dots~\cite{akopian:103501, ghali:661}.
However, polarization encoding is prone to dispersion in optical fibers, which 
affect the polarization of the outcoupled photons. Thus, the polarization 
encoding is not well-suited for large distance quantum communication in
real-world implementations~\cite{antonelli:080404}. 
An alternative is time-bin entangled light states~\cite{marcikic:062308}, 
where 
quantum information is encoded in the arrival time of photons. 
Time-bin 
entangled states are more robust to the decoherence in optical 
fibers~\cite{thew:062304} enabling distribution of entangled photons 
over 300 km \cite{inagaki:23241}. In these experiments spontaneous parametric 
down-conversion is used as a source 
of time-bin entangled photons~\cite{brendel:2594, zavatta:020502}. However, 
parametric 
down-conversion is a random process with probabilistic number of generated 
photon. The generation of multiple pairs reduces the accuracy and security of 
the quantum communication. Alternatively, schemes for generation of single 
pairs of 
time-bin entangled photons using the biexciton cascade in a quantum
dot have been proposed~\cite{simon:030502}. 
These schemes require the preparation of the quantum dot into a long-lived or 
metastable state, which is non-trivial challenge, that prevents an experimental 
implementation of such schemes.

In the present article,
we show a scheme for generation a single-photon qubit time-bin-encoded across 
superposed two spatiotemporal peaks. In an earlier work, within the frame of 
exact cavity quantum electrodynamics~\cite{khanbekyan:013822}, we have studied 
the theory of 
generation of single-photon wave packets by means of interaction of the 
cavity-assisted quantized electromagnetic field with a pumped three-level 
$\Lambda$-type emitter~\cite{khanbekyan:013803}. In particular, the possibility 
to generate single-photon wave packets with requested spatiotemporal shapes has 
been shown. Here, we study the generation of photon wave packets of double-peak 
shapes---a possible implementation of time-bin entanglement. 
In particular, we show that by adjusting the
the driving laser pulse 
modulation of the phase difference between the spatiotime
bins of the generated single-photon state can be achieved.

%%%%%%%%%%%%%%%%%%%%%%%%%%%%%%%%%%%%%%%%%%%%%%%%%%%%%%%%%%%%%%%

\section{Basic equations}
\label{sec2}

\begin{figure}[t!]
	\begin{center}
	\includegraphics[width=0.48\textwidth]{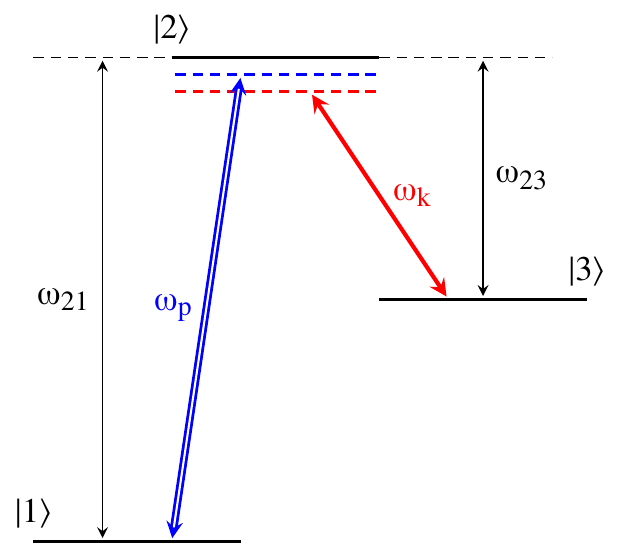}
	\caption{
		\label{fig1}
		Scheme of energy levels and transitions for the three-level
		$\Lambda$-type atomic emitter.}
		\end{center}
\end{figure}

We consider a single atom-like emitter 
that interacts with the electromagnetic field
in the presence of a dispersing and absorbing dielectric medium with a
spatially
varying and frequency-dependent complex permittivity.
We assume that only a single transition
(\mbox{\ket{2} $\!\leftrightarrow$ \!\ket{3}}, frequency $\omega_{23}$)
is quasi resonantly
coupled to a narrow-band cavity-assisted electromagnetic field
(frequency $\omega_k$), cf. Fig.~\ref{fig1},
and that an external (classical) pump field with
quasi resonant frequency $\omega_p$ and
(time-dependent)
Rabi frequency
$\Omega_p(t)$ is applied to the
\mbox{\ket{1} $\!\leftrightarrow$ \!\ket{2}}
transition (frequency $\omega_{21}$), cf. Fig.~\ref{fig1}.
For the sake of simplicity of presentation, we restrict our treatment to
the one-dimensional case ($z$ axis) and assume that the resonator cavity
is formed by an empty body
bounded
by an outcoupling fractionally transparent mirror at $z=0$ and
a perfectly reflecting mirror at $z=-l$,
and (negative) $z_A$ is the position of the emitter
inside the cavity.
Applying the multipolar-coupling scheme in electric dipole
approximation and the
rotating-wave approximation, we may write the Hamiltonian
that governs the temporal evolution
of the overall system, which consists of the electromagnetic
field, the dielectric medium (including the dissipative degrees
of freedom), and the emitter coupled to the field,
in the form of (for details, see Refs.~\cite{khanbekyan:013822, 
khanbekyan:013803})
\begin{align}
\label{1.15}
\hat{H} =
&
\int\! \D z\int_0^\infty\! \D\omega
\hbar\omega\hat {f}^{\dagger}(z, \omega)
\hat{f}(z, \omega)
+ \hbar \omega _{21} \hat {S}_{22} + \hbar \omega _{31} \hat {S}_{33}
\nonumber\\[1ex]&
-
\frac{\hbar}{2}
\Omega_p
(t)\left[
\hat{S}_{12}^{\dagger}
e^{-i\omega_p t}
+
\mbox{H.c.} \right]
-g(t)\left[
d_{23}\hat{S}_{32}^{\dagger}
\hat{E}^{(+)}(z_A)
+
\mbox{H.c.} \right]
,
\end{align}
with \mbox{$\omega _{31} $ $\!=$ $\omega _{21}- \omega _{23}$}.
In this equation, the first term is the Hamiltonian of
the \mbox{field--me}\-dium system, where the
fundamental
bosonic fields
\mbox{$f(z,\omega)$}
and \mbox{$f^\dagger(z,\omega)$},
\begin{align}
\label{1.3}
&      \bigl[f (z, \omega),
f^{\dagger } (z',  \omega ') \bigr]
= 
\delta (\omega - \omega  ')
\delta
(z - z') ,
\\
\label{1.3-1}
&\bigl[f (z, \omega),
f (z',  \omega ') \bigr]
= 0,
\end{align}
play the role of the canonically conjugate system variables.
The second and the third terms represent the Hamiltonian of the emitter, where
the $\hat{S}_{kk'}$ are the flip operators,
\begin{equation}
\label{1.5}
\hat{S} _{kk'} =
| k\rangle
\!\langle k' |
,
\end{equation}
corresponding to the \mbox{\ket{k} $\!\leftrightarrow$ \!\ket{k'}} transition
with the frequency $\omega_{kk'}$,
where
$|k\rangle$
being the energy eigenstates
of the emitter.
Finally, the fourth term is the emitter--pump coupling energy and the last  
term is the emitter--field
coupling energy, where
\begin{equation}
\label{1.7}
\hat{ d}_A = \sum _{kk'}
d _{Akk'}  \hat{S} _{kk'}
\end{equation}
is the electric dipole-moment operator
($ d _{Akk'}$ $\!=$ $\!\langle k|
\hat{ d}_{\!A} | k' \rangle$),
and 
the (real) time-dependent
function $g(t)$ defines the (time-dependent)
shape
of the
interaction of the emitter with the cavity field, which without loss
of generality can be
chosen to be
normalized to unity.
The operator of the
medium-assisted electric field $\hat{E}(z)$
can be expressed in terms of the variables
$\hat{f}(z,\omega)$ and
$\hat{f}^\dagger(z,\omega)$ as
follows:
\begin{equation}
\label{1.9}
\hat{E}(z) = \hat{E}^{(+)}(z)
+\hat{E}^{(-)}(z),
\end{equation}
\begin{equation}
\label{1.10}
\hat{E}^{(+)}(z) = \int_0^\infty \D\omega\,
\uh{E}(z,\omega),
\quad
\hat{E}^{(-)}(z) =
[\hat{E}^{(+)}(z)]^\dagger,
\end{equation}
\begin{align}
\label{1.11}
 \uh{E}(z,\omega) =
i \sqrt{\frac {\hbar}{\varepsilon_0\pi}}\,
\frac{ \omega^2}{c^2}
\int \D^3r'\sqrt{\varepsilon''(z',\omega)}\,
G(z,z',\omega)
%\cdot
\hat{f}(z',\omega).
\end{align}
%In the above, the classical (retarded)
%Green tensor $G(z,z',\omega)$
%is the solution to the equation
%\begin{equation}
%\label{1.13}
%\nabla
%\times
%\nabla\!  \times G  (z, z', \omega)
%- \frac {\omega ^2 } {c^2} \,\varepsilon (z ,\omega)
%G  (z, z', \omega)
%=  \delta (z-z')
%\end{equation}
%together with the boundary condition at infinity,
%\mbox{$G(z,z',\omega)\to 0$} if
%\mbox{$|z-z'|\to\infty$}, 
In the above, $G(z,z',\omega)$ is the classical (retarded) Green tensor for the 
Helmholtz equation, that
defines the structure of the
electromagnetic field
formed by
the present dielectric bodies.

In what follows we assume that
the emitter is initially (at time $t$ $\!=$ $\!0$)
prepared in the state \ket{1} and the rest of
the system, i.\,e., the part of the
system that consists of the
electromagnetic field and the dielectric media
(i.\,e., the cavity),
is prepared in the ground
state \ket{\{0\}}.
Since, in the case under consideration,
we may approximately span the Hilbert space of the whole system by the
single-excitation states,
we expand the state
vector of the overall system at later times $t$ ($t\ge 0$) as
\begin{align}
\label{1.17}
&
|\psi(t)\rangle =
C_1(t)
|\!\left\lbrace0\right\rbrace\!\rangle|1\rangle +
C_2(t)
e^{-i\omega_{21} t}
|\!\left\lbrace0\right\rbrace\!\rangle |2\rangle
\nonumber\\[1ex]
&
+\!\int\! \D z \int_0^\infty \!\D \omega\,
C_3(z, \omega, t)
e^{-i(\omega +\omega _{31})t}
\hat{f}^{\dagger}(z, \omega)
|\!\left\lbrace0\right\rbrace\!\rangle |3\rangle ,
\end{align}
where $\hat{f}^{\dagger}(z, \omega)|\{0\}\rangle$ is
an excited single-quantum state of the combined
field--cavity system.

It is not difficult to prove
that the Schr\"odinger equation for
\ket{\psi(t)}
leads to the following system of
differential
equations for the probability amplitudes
$C_1(t)$, $C_2(t)$ and $C_3(z, \omega, t)$:
\begin{equation}
\label{1.19}
\dot {C_1} =
\frac{i}{2}
\Omega_p
(t)
e^{i\Delta_pt}
C_2(t),
\end{equation}
\begin{multline}
\label{1.20}
\dot {C_2} =
\frac{i}{2}
\Omega_p
(t)
e^{-i\Delta_pt}
C_1(t)
%\\[.5ex]
-\frac{d_{23}}{\sqrt{\pi \hbar \varepsilon _0  \mathcal{A}}}
\int_0^\infty\! \D\omega\, \frac{\omega ^2}{c^2}
\int \D z
\sqrt{\varepsilon''(z,\omega)}\,
\\[.5ex]
\times
G(z_A, z,\omega)
C_3(z, \omega, t)g(t)
e^{-i (\omega - \omega_{23})t},
\end{multline}
\begin{multline}
\label{1.18}
\dot {C_3}(z, \omega, t) =
\frac{d_{23}^*}{\sqrt{\pi \hbar \varepsilon _0  \mathcal{A}}}
\frac{\omega ^2}{c^2}
\sqrt{\varepsilon''(z,\omega)}\,
%\\[.5ex]
%\times
G^*(z_A, z,\omega)
C_2(t)g(t)
e^{i (\omega - \omega_{23})t},
\end{multline}
where $\mathcal{A}$ is the area
of the coupling mirror of the cavity,
and $\Delta_p=\omega_p-\omega_{21}$
is the detuning of
the pump frequency
from the \mbox{\ket{1} $\!\leftrightarrow$ \!\ket{2}} transition
frequency.
The Green function $G(z,z',\omega)$ determines the spectral response
of the resonator cavity.
In particular, its poles define quasidiscrete set of lines, where we assume 
that the $k$th mode of the cavity with the complex frequency
\begin{equation}
\label{1.21}
\tilde{\omega}_k
= \omega_{k}
- {\textstyle\frac {1} {2}}i\Gamma _{k},
\end{equation}
is quasi-resonantly coupled to the transition
\mbox{\ket{2} $\!\leftrightarrow$ \!\ket{3}} with the transition
frequency $\omega_{23}$
(cf. Fig.~\ref{fig1}).
Then, 
substituting the formal solutions to Eqs.~(\ref{1.19}) and (\ref{1.18})
[with the initial condition \mbox{$C_1(0)=1$}, \mbox{$C_2(0)=0$} and
\mbox{$C_3(z, \omega, 0)=0$}] into  Eq.~(\ref{1.20}),
we can
derive the integro-differential equation
\begin{align}
\label{1.22}
\dot {C_2} =
\frac{i}{2}
\Omega
_p(t)
e^{-i\Delta_pt}
+
\int_0^t \! \D t'\,
K(t,t')
C_2(t'),
\end{align}
where the kernel function $K(t,t')$ reads
\begin{align}
\label{1.24}
K(t,t')=
-\frac{1}{4}
\Omega
_p(t)
\Omega
_p(t')
e^{-i\Delta_p(t-t')}
%\nonumber\\[1ex]&
-
\frac{1}{4}
\alpha_k 
\tilde{\omega}_k
g(t)
g(t')
e^{-i(\Delta_k - i \Gamma_k/2)(t-t')},
\end{align}
with \mbox{$\Delta_k$ $\!=\omega_k$ $\!-\omega_{23}$} and
\begin{equation}
\label{1.23}
\alpha_k  =  \frac{4|d_{23}|^2 }
{\hbar \varepsilon _0  \mathcal{A}
	l}\,
\sin^2
(\omega_k
z_A/c)
.
\end{equation}
From Eq.~(\ref{1.22}) together with Eq.~(\ref{1.24})
we can conclude
that
\mbox{$R_k$ $\!\equiv\sqrt{\alpha_k\omega_k}$}
can be
regarded as vacuum Rabi frequency of emitter--cavity interaction.

%%%%%%%%%%%%%%%%%%%%%%%%%%%%%%%%%%%%%%%%%%%%%%%%%%%%%%%%%%%%%%%
\section{Single-photon Generation and wave-packet shape}
\label{sec5}
%%%%%%%%%%%%%%%%%%%%%%%%%%%%%%%%%%%%%%%%%%%%%%%%%%%%%%%%%%%%%%%

Following Ref.~\cite{khanbekyan:013822}, it can be shown that
when the Hilbert space of the system is
effectively spanned by a single excitation,
%on a time-scale that is short compared to the inverse spontaneous
%emission rate of the emitter,
the Wigner function of the quantum state
of the excited outgoing wave packet can be derived to be 
\begin{equation}
\label{1.27}
W(\alpha,t)
= [1-\eta(t)]W^{(0)}(\alpha)
+\eta(t)W^{(1)}(\alpha),
\end{equation}
with $W^{(0)}(\alpha)$ and $W^{(1)}(\alpha)$
being the Wigner functions of the vacuum
state and the one-photon Fock state. As we see,
the excited outgoing mode
is basically prepared
in a mixed state of a one-photon Fock state and the
vacuum state, due to
unavoidable existence of unwanted losses, where
$\eta(t)$ can be regarded as being the efficiency
to prepare the excited outgoing wave packet in
a one-photon Fock state:
\begin{equation}
\label{1.29}
\eta(t)
=
\int_0^{\infty}\!
\D\omega\, |F(\omega ,t)| ^2
\simeq
\int_{-\infty}^\infty
\D\omega\, |F(\omega ,t)| ^2
,
\end{equation}
with
\begin{multline}
\label{1.30}
F(\omega, t)=
\frac{d_{23}}{\sqrt{\pi \hbar \varepsilon _0  \mathcal{A}}}
\sqrt{\frac{c}{\omega}}
\frac{\omega^2}{c^2}
\\[1ex]\!\!\!
\times\!
\!\!      \int ^t _0\! \D t'
G^*(0^+, z_A, \omega)
C_2^*(t')g(t') e^{i\omega(t-t')}
e^{i
	\omega
	_{23}t'}
e^{i\omega_{31}t}
,\!\!\!\!
\end{multline}
where $0^+$ indicates the position \mbox{$z$ $\!=0$} outside the cavity.
The excited outgoing wave packet
(mid-frequency $\omega_k$)
is characterized by the mode function
\begin{equation}
\label{1.31}
F_1(\omega,t)
= \frac{F(\omega , t)}{\sqrt{\eta(t)}}\,
\end{equation}
and spatiotemporal shape 
\begin{equation}
\label{1.35}
\phi(z, t)
=
{\textstyle\frac{1}{2}}
\int_0^{\infty}\!
\D\omega\,
\sqrt{\frac{\hbar\omega}{\varepsilon_0 c \pi\mathcal{A}
}}\,
e^{-i \omega z/c}
F_1(\omega, t)
.
\end{equation}
Finally, inserting
Eq.~(\ref{1.31}) together with Eq.~(\ref{1.30}) into Eq.~(\ref{1.35}),
we find the spatiotemporal shape of the excited outgoing wave packet
outside the cavity as
(in the following, for the sake of simplicity, we assume \mbox{$\omega_{31}$ 
$\!=0$})
\begin{multline}
\label{2.1}
\phi(z,t)=
\frac{ R _k}{2}
\sqrt{
	{\displaystyle
		\frac{\hbar\omega_k \gamma_{k\mathrm{rad}}}
		{2\varepsilon_0 c\mathcal{A} \eta(t)}
}}
%\\[.5ex] \times\!
\int_0^{t-z/c}
\!\D  t'\,
C_2^*(t')g(t')
e^{-i(\Delta_k + i \Gamma_k/2) t'}
e^{i 
	\tilde{\omega}_k^*(t-z/c)
}
,
\end{multline}
where $\gamma_{k\mathrm{rad}}$ describes the wanted radiative
losses due to transmission of the radiation through the fractionally
transparent mirror.

%%%%%%%%%%%%%%%%%%%%%%%%%%%%%%%%%%%%%%%%%%%%%%%%%%%%%%%%%%%%%%%
\section{Generation of time-bin wave packets}
\label{sec7}
%%%%%%%%%%%%%%%%%%%%%%%%%%%%%%%%%%%%%%%%%%%%%%%%%%%%%%%%%%%%%%%

As we can see from Eq.~(\ref{2.1}) together with Eq.~(\ref{1.22})  the shape of 
single-photon outgoing wave packet can be changed by changing the
rates and shapes of the  
emitter-cavity interaction and/or the driving pulse. In the context of 
time-bin entangled states of light, a question of interest is the generation  
of single-photon wave packets with a double-peak spatiotemporal structure, 
where every of these 
peaks have  probability to carry the photon. 
Thus, let us assume that the desired double-peak spatiotemporal shape of the
single-photon outgoing
wave packet is given by a function $\phi(z,T)$ at
some
time $T$, where
the condition
\mbox{$T$ $\!\gg$ $\!\Gamma_k^{-1}$}
ensures
that the wave packet
has
almost completely
left
the cavity.
From
Eq.~(\ref{2.1})
we find that
\begin{align}
\label{7.1}
C_2(t) &=
\frac{2}{R_kg(t)} \sqrt{\frac{2\varepsilon_0c
		\mathcal{A}\eta(T)}{\hbar \omega_k \gamma_{k\mathrm{rad}}}}
e^{-i \omega_{23} t}
%\nonumber\\[1ex]&\times
\left\lbrace
\frac{\D \phi[c(T-t),T]}{\D t}
-i 
\tilde{\omega}_k^*
\phi[c(T-t),T]
\right\rbrace
.
\end{align}
Further, inserting Eq.~(\ref{1.24}) into Eq.~(\ref{1.22}) we see that
\begin{align}
\label{7.3}
D(t) = f(t) + \int_0^t \D t' f(t)f(t')C_2(t'),
\end{align}
where $D(t)$ is defined by
\begin{align}
\label{7.5}
D(t) \equiv \dot{C}_2(t)
+ \frac{R_k^2}{4} \int _0^t \D t' C_2(t')
g(t)g(t')
e^{-i(\Delta_k - i \Gamma_k/2)(t-t')},
\end{align}
and $f(t)$ is related to the
shape
of the pump pulse $\Omega_p(t)$ according to
\begin{align}
\label{7.7}
f(t) = \frac{i}{2} \Omega_p(t) e^{-i\Delta_pt}.
\end{align}
Differentiation of Eq.~(\ref{7.3}) with respect to $t$ then yields the
following differential equation for $f(t)$:
\begin{align}
\label{7.9}
D(t)\dot{f}(t) + C_2(t) f^3(t)-\dot{D}(t)f(t)=0 .
\end{align}
Hence,
for the desired double-peak spatio-temporal shape $\phi(z,T)$
of the single-photon outgoing wave packet, the solution of the 
differential
equation~(\ref{7.9}) yields the sought
shape
of the driving pulse.

% % % % % % % % % % % % % % % % % % % % % % % % % % % % % % % %
\begin{figure}[t!]
	\includegraphics[width=0.5\textwidth]{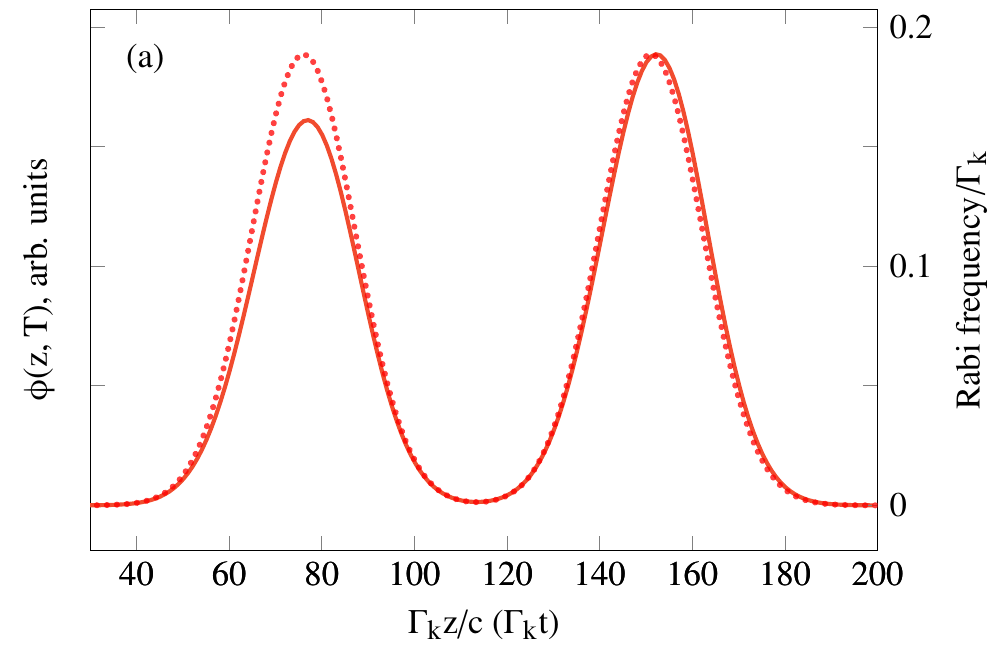}\hspace{2ex}
	\includegraphics[width=0.5\textwidth]{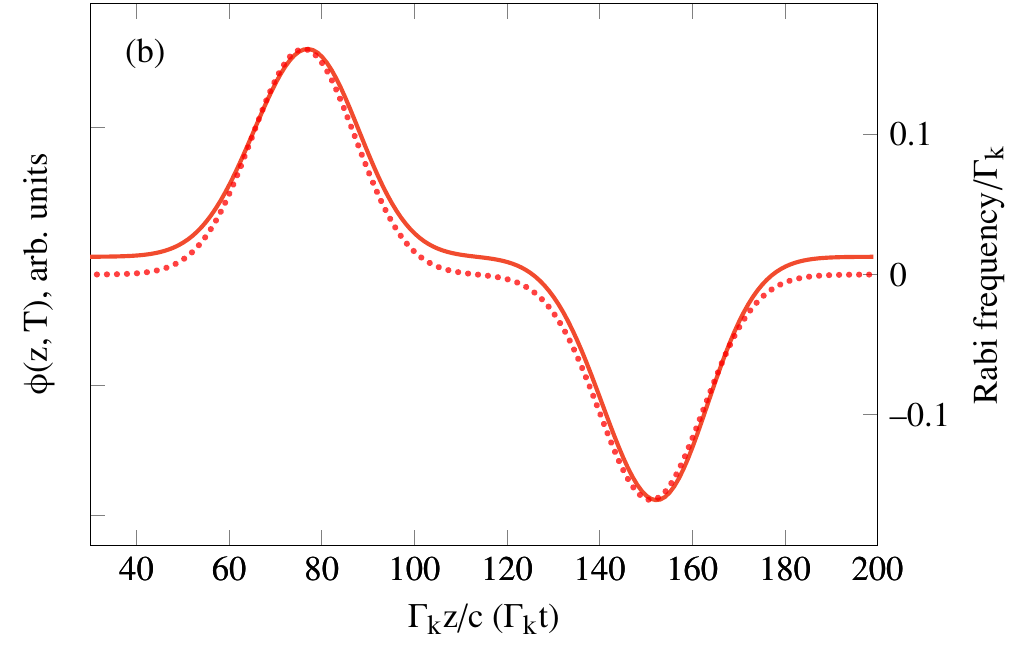}
	\caption{
		Driving pulse (solid lines) required to generate double-peak wave 
		packets (dotted lines) of the
		one-photon Fock state
		outgoing field 
		for
		(a) the symmetric case and (b) the case with $\pi$-phase between the 
		peaks with
		$R_k = 2 \Gamma_k$, 
		$\Omega_{p\mathrm{max}} = 0.7\Gamma_k$, $\Delta_k = \Delta_p = 0$, $T = 
		200 \Gamma_k$ and efficiency $\eta(T)=0.9$.}
	\label{fig5}
\end{figure}
% % % % % % % % % % % % % % % % % % % % % % % % % % % % % % % %
In Fig.~\ref{fig5}(a) we illustrate the shape of the driving pulse required for 
the generation of a symmetric double-peak wave packet in the case, when the 
emitter-cavity interaction is constant. Assuming equal 
probabilities for finding the photon in every of the time bins, this 
corresponds to a quantum superposition state of the form 
\mbox{\ket{\Psi} $\!=$ $\textstyle{\frac{1}{\sqrt{2}}}$ 
$\!(\left|01\right\rangle$ $\!+$ $\!\left|10\right\rangle)$}, where \ket{01} 
and \ket{10} describe finding the photon in the first and second bin, 
accordingly. In Fig.~\ref{fig5}(b) we present the case,  when a phase 
difference equal $\pi$ between the first and the second peaks is established.

\begin{figure}[t!]
	\begin{center}
	\includegraphics[width=0.75\textwidth]{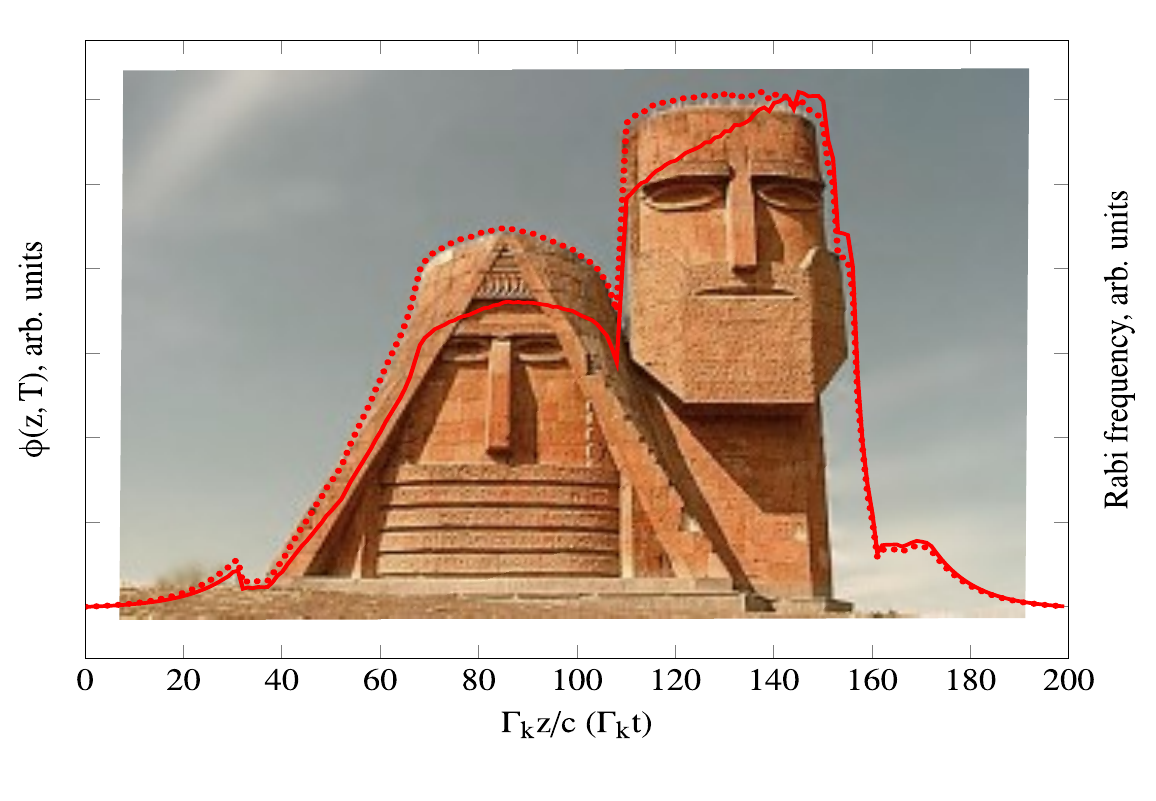}
	\end{center}
 	\caption{
          Shape
          of the driving pulse (solid red line) required to generate
          single-photon outgoing wave packets whose shape (dotted red
          line) resembles the "Tatik-Papik" monument.}
	\label{fig9}
\end{figure}

Finally, the proposed method allows the generation of double-peak 
wavepackets of even more complex structure. Let us have a look at  
Fig.~\ref{fig9}, which illustrates the generation of
an asymmetric two-peak single-photon wave packet the shape of which
resembles the monument called "Tatik-Papik" ("Grandma-Grandpa").

In summary, 
we have proposed a scheme for generation of single-photon time-bin double-peak 
wave packets based on resonant interaction of a three-level $\Lambda$-type 
emitter in a high-$Q$ cavity. We have shown, that control over relative phase 
of the individual time-bins is possible. Importantly, the results illustrate 
the ability to encode large amount of information within a single photon.

\bibliographystyle{ieeetr}

%\bibliography{bibl}

\begin{thebibliography}{10}

\bibitem{gisin:165}
{\bf N.~Gisin and R.~Thew}.  {\em Nat. Photonics}, {\bf 1}, 165 (2007).

\bibitem{knill:46}
{\bf E.~Knill, R.~Laflamme, and G.~J. Milburn}.  {\em Nature}, {\bf 409}, 46
  (2001).

\bibitem{bouwmeester:575}
{\bf D.~Bouwmeester, J.-W. Pan, K.~Mattle, M.~Eibl, H.~Weinfurter, and
  A.~Zeilinger}.  {\em Nature}, {\bf 390}, 575 (1997).

\bibitem{riedmatten:050302}
{\bf H.~de~Riedmatten, I.~Marcikic, J.~A.~W. van Houwelingen, W.~Tittel,
  H.~Zbinden, and N.~Gisin}.  {\em Phys. Rev. A}, {\bf 71}, 050302 (2005).

\bibitem{jennewein:4729}
{\bf T.~Jennewein, C.~Simon, G.~Weihs, H.~Weinfurter, and A.~Zeilinger}.  {\em
  Phys. Rev. Lett.}, {\bf 84}, 4729 (2000).

\bibitem{deng:042317}
{\bf F.-G. Deng, G.~L. Long, and X.-S. Liu}.  {\em Phys. Rev. A}, {\bf 68},
  042317 (2003).

\bibitem{kwiat:4337}
{\bf P.~G. Kwiat, K.~Mattle, H.~Weinfurter, A.~Zeilinger, A.~V. Sergienko, and
  Y.~Shih}.  {\em Phys. Rev. Lett.}, {\bf 75}, 4337 (1995).

\bibitem{akopian:103501}
{\bf N.~Akopian, N.~H. Lindner, E.~Poem, Y.~Berlatzky, D.~G. J.~Avron, B.~D.
  Gerardot, and P.~M. Petroff}.  {\em Phys. Rev. Lett.}, {\bf 96}, 103501
  (2006).

\bibitem{ghali:661}
{\bf M.~Ghali, K.~Ohtani, Y.~Ohno, and H.~Ohno}.  {\em Nat. Comm.}, {\bf 3},
  661 (2012).

\bibitem{antonelli:080404}
{\bf C.~Antonelli, M.~Shtaif, and M.~Brodsky}.  {\em Phys. Rev. Lett.}, {\bf
  106}, 080404 (2011).

\bibitem{marcikic:062308}
{\bf I.~Marcikic, H.~de~Riedmatten, W.~Tittel, V.~Scarani, H.~Zbinden, and
  N.~Gisin}.  {\em Phys. Rev. A}, {\bf 66}, 062308 (2002).

\bibitem{thew:062304}
{\bf R.~T. Thew, S.~Tanzilli, W.~Tittel, H.~Zbinden, and N.~Gisin}.  {\em Phys.
  Rev. A}, {\bf 66}, 062304 (2002).

\bibitem{inagaki:23241}
{\bf T.~Inagaki, N.~Matsuda, O.~Tadanaga, M.~Asobe, and H.~Takesue}.  {\em Opt.
  Exp.}, {\bf 21}, 23241 (2013).

\bibitem{brendel:2594}
{\bf J.~Brendel, N.~Gisin, W.~Tittel, and H.~Zbinden}.  {\em Phys. Rev. Lett.},
  {\bf 82}, 2594 (1999).

\bibitem{zavatta:020502}
{\bf A.~Zavatta, M.~D'Angelo, V.~Parigi, and M.~Bellini}.  {\em Phys. Rev.
  Lett.}, {\bf 96}, 020502 (2006).

\bibitem{simon:030502}
{\bf C.~Simon and J.-P. Poizat}.  {\em Phys. Rev. Lett.}, {\bf 94},
  030502 (2005).

\bibitem{khanbekyan:013822}
{\bf M.~Khanbekyan, D.-G. Welsch, C.~Di~Fidio, and W.~Vogel}.  {\em Phys.\ Rev.\
  A}, {\bf 78}, 013822 (2008).

\bibitem{khanbekyan:013803}
{\bf M.~Khanbekyan and D.-G. Welsch}.  {\em Phys. Rev. A}, {\bf 95}, 013803
  (2017).

\end{thebibliography}

\end{document}